\documentclass[11pt]{article}
\usepackage{epsfig}

\begin{document}
\def\beq{\begin{equation}}
\def\eeq{\end{equation}}
\vskip 30pt

\begin{center}
\rightline{September 2001}
\rightline{UCOFIS 4/01}
\rightline{US-FT/10-01}
\vskip 0.5cm
{\large \bf NUCLEAR STRUCTURE FUNCTIONS AND HEAVY FLAVOUR LEPTOPRODUCTION
OFF THE 
NUCLEUS AT SMALL $X$ IN PERTURBATIVE QCD}
\vskip 20pt
N. Armesto\\
{\it Departamento de F\'{\i}sica, M\'odulo C2, Planta baja, Campus de
Rabanales,}\\ {\it Universidad de C\'ordoba, E-14071 C\'ordoba, Spain}\\
\vskip 0.2cm
and\\
\vskip 0.2cm
M. A.
Braun\footnote{On leave of absence from
Department of High-Energy Physics, St. Petersburg
University, 198504 St. Petersburg, Russia.}
\\ {\it Departamento de F\'{\i}sica de Part\'{\i}culas, Universidade de
Santiago de Compostela,}\\ {\it E-15706 Santiago de Compostela, Spain}
\end{center}

\vskip 30pt
\begin{abstract}

Nuclear structure functions and cross-sections for heavy flavour
production in lepton-nucleus collisions are investigated in the
low $x$ region accessible now or in the near future. The
scattering on a heavy nucleus
is described by the sum of fan
diagrams of BFKL pomerons, 
which is exact in the high-colour limit. The initial condition for the 
evolution at $x=0.01$ is taken from a saturation model, which 
reproduces the
experimental data on the proton.
The $A$ dependence of the  structure functions is well described by
a power factor $A^\alpha$, with
$\alpha$ reaching values as low as 1/2 at extremely low $x$. The total 
cross-sections for heavy 
flavour production reach values of the order of mb, and the corresponding
transverse momentum
distributions are sizeable up to transverse momenta larger than the
initial large scale $\sqrt{Q^2+4m_f^2}$.

\end{abstract}

\section{Introduction}

In 
the framework of the colour dipole model [1,2] and
in the high-colour limit $N_c\to\infty$, the scattering on a heavy nucleus 
is exactly described by the sum of fan
diagrams constructed of BFKL pomerons, each of them splitting into two.
Numerical solutions of the
resulting equation for the colour dipole cross-section on the
nucleus [3-9] were first presented in [8], and then in [10-14].
The gluon density 
introduced in
[8] revealed a "supersaturation" behaviour, tending to zero at any fixed
momentum $k$ as the rapidity $Y\to\infty$. As a function of $\ln k$ it proved
to have a form of a soliton wave moving to the right with a constant
velocity as $Y$ increases. This latter phenomenon is related to its scaling 
property: at high enough $Y$ the gluon density at fixed impact 
parameter $b$ proved to be a function
of the ratio $k/Q_s(Y,b)$, where $Q_s(Y,b)$, the position in momentum
space of the maximum of the gluon density,
can be interpreted as a "saturation momentum" growing as a power of energy.
Subsequent numerical calculations [10,14] confirmed this scaling
behaviour,
although the gluon density itself is defined differently by
different authors.

Obviously partonic densities in the nucleus are not observable themselves
but are only certain theoretical tools to calculate the observable 
quantities. Therefore different definitions of the gluonic density
are well admissible, provided they lead to the same observables.
As such the structure function of the nucleus is the most directly
calculable one. 
It was first calculated in [8] from the found solution
of the BFKL fan diagram equation for a very wide range of energies
(up to rapidities $\sim 50$), for the nucleus in the form of a
spherical
well and for a rather specific colour distribution inside the nucleon,
not adapted to the experimental data. The found structure function
proved to be rising as $Y$ with rapidity and roughly linearly with
$Q^2$ in agreement with the theoretical expectations [7,11].
Later calculations with a more realistic nuclear density [15] revealed that
the nuclear structure function in fact grows  as $Y^2$, the extra growth
provided by the contribution of peripheral collisions not damped by
the non-linear term in the evolution equation. Still the attention in 
both [8] and [15]
was centered at the asymptotic behaviour at high energies, insensitive
to the choice of the initial conditions and governed only by the
internal dynamics of the evolution equation.

However, from the experimental point of view it is desirable to know
the structure function at values of $x$  accessible now or in the near future
(see e.g. [16]), say down to $10^{-7}$ corresponding to $Y\leq 16\div 17$.
At such rapidities the influence of the initial condition is still rather
noticeable so that its choice becomes a matter of importance.
Accordingly we performed a new run of calculations employing
more realistic colour distributions in the proton, following from
the saturation approach of [17] consistent with DIS data for the proton 
below $x=0.01$\footnote{Other approaches [18] which explicitly contain
saturation in the form of multiple scattering and also describe the
experimental data for the proton, could be used, see [19] for a study of
saturation in the low $Q^2$ region within this multiple scattering
model. We will restrict ourselves to the use of [17] due to technical
reasons, see Section 3.}.
In [10] the parton distributions following from this
initial condition were studied. In this paper we  report on the 
nuclear structure
function following from the BFKL fan diagram evolution equation
with physically supported initial conditions. 
The solution of this equation also allows us to find the gluon 
density
in the nucleus. We use it to study another directly measurable quantity:
the inclusive probability for heavy flavour production off nuclei (by
photon-gluon fusion) in photon
induced reactions.

\section{The nuclear structure function from the evolution
equation}

In this section we collect the formulae necessary to calculate the 
structure function from the solution of the BFKL fan diagram
evolution equation.
The nuclear structure function $F_2$  can be standardly defined via
the cross-sections $\sigma_{T,L}$  for the collision of 
the transversal ($T$) or longitudinal ($L$) virtual
photon of momentum $q$, $q^2=-Q^2$, on the nucleus $A$ of momentum $Ap$:
\beq
F_2(x,Q^2)=\frac{Q^2}{\pi e^2}(\sigma_T+\sigma_L), \label{eq1}
\eeq
where $x=\exp(-Y)=Q^2/(2q\cdot p)$.
Both cross-sections 
can be conveniently presented via the cross-section 
$\sigma_{dA}(Y,r)$ for the scattering of a colour dipole of the transverse 
radius $r$ on the nucleus:
\beq
\sigma_{T,L}(Y)=\int d^2r\rho_{T,L}(r)\sigma_{dA}(Y,r), \label{eq2}
\eeq
where $\rho_{T,L}$ is a well-known distribution of colour dipoles
created by splitting of the incident photon into $q\bar q$ pairs:
\beq
\rho_T(r)= \frac{e^{2}N_c}{8\pi^3}\sum_{f}Z_f^2
\int_{0}^{1}d\alpha
\Big\{[\alpha^{2}+(1-\alpha)^{2}]\epsilon^{2}{\rm K}_{1}^{2}(\epsilon r)+
m_f^2{\rm K}_0^2(\epsilon r)\Big\} \label{eq3}
\eeq
and 
\beq
\rho_{L}(r)= \frac{e^{2}N_c}{2\pi^3}Q^2\sum_fZ_f^2
\int_{0}^{1}d\alpha
\alpha^2(1-\alpha)^2{\rm K}_{0}^{2}(\epsilon r). \label{eq4}
\eeq
Here summation goes over flavours,
$\epsilon^2=Q^{2}\alpha (1-\alpha)+m_f^2$, and $m_f$ and $Z_f$ are
respectively the mass and 
electric charge in units of $e$, of the quark of flavour $f$.

The dipole-nucleus cross-section in its turn can be presented as an integral 
over the impact parameter $b$:
\beq
\sigma_{dA}(Y,r)=2\int d^2b\ \Phi(Y,r,b), \label{eq5}
\eeq
where  $2\Phi$ has a meaning of the cross-section at fixed
impact parameter. The evolution equation in $Y$ can be most conveniently
written for the function
\beq
\phi(Y,r,b)=\frac{1}{2\pi r^2}\Phi(Y,r,b) \label{eq6}
\eeq
in momentum space. It reads [8]
\beq
\left(\frac{\partial}{\partial y}+H_{BFKL}\right)\phi(y,q,b)=
-\phi^2(y,q,b), \label{eq7}
\eeq
where  $y=\alpha_sN_cY/\pi$ and 
$H_{BFKL}$ is the forward BFKL Hamiltonian.

Putting (\ref{eq6}) into (\ref{eq5}) and (\ref{eq2}),
and passing to the momentum space, we can express
the cross-sections $\sigma_{T,L}$ directly via the function $\phi$ or
its 2nd derivative in the logarithm of the momentum:
\beq
\sigma_{T,L}=\frac{1}{\pi} \int d^2b
d^2q\ \phi(y,q,b)w_{T,L}(q)=
\frac{1}{\pi}\int d^2b\frac{d^2q}{q^2}h(y,q,b)\tilde{\rho}_{T,L}(q),
\label{eq8}
\eeq
where 
\beq
\tilde{\rho}_{T,L}(q)=\int d^2r\ \rho_{T,L}(r)\left(1-e^{i{\bf qr}}\right),\ \ 
w_{T,L}(q)=\int d^2r\  r^2\rho_{T,L}(r)e^{i{\bf qr}}  \label{eq9}
\eeq
and 
\beq
h(y,q,b)=q^2\nabla_q^2\phi(y,q,b)=\frac{\partial^2\phi(y,q,b)}
{(\partial\ln q)^2}\ .  \label{eq10}
\eeq
Up to a trivial factor, function $h$ gives the gluon density in the 
nucleus [8]:
\beq
\frac{\partial xG(x,k^2,b)}{\partial^2 b \partial k^2}=
\frac{2 N_c}{\pi
g^2}k^2\nabla_k^2\phi\left(\ln\frac{1}{x},k,b\right).
\label{eq11}
\eeq
This definition is not unique (see the discussion in [20]) but natural
in the sense that the structure function can
be directly expressed via the gluon density thus defined.
We shall see later that this is also true  for other observables.

Note that for numerical calculations the first form in (\ref{eq8})  is 
much more
convenient, since it requires neither numerical differentiation
nor taking into account the singularity at $q=0$. Functions $w_{T,L}$ can be
easily found analytically:
\beq
w_T(q)= \frac{e^{2}N_c}{8\pi^2}\nabla_q^2\sum_fZ_f^2
\int_{0}^{1}d\alpha
\Big[\left(\alpha^{2}+(1-\alpha)^{2}\right)(q^2/2
+\epsilon^2)-m_f^2\Big]J(q,\epsilon),  \label{eq12}
\eeq
where
\beq
J(q,\epsilon)=
\frac{2}{q\sqrt{q^2+4\epsilon^2}}\ln\frac{\sqrt{q^2+4\epsilon^2}+q}
{\sqrt{q^2+4\epsilon^2}-q}\ .  \label{eq13}
\eeq
The longitudinal density $w_{L}$ is given by the same expression
without the term with $m_f^2$ and 
with the substitutions
\[\alpha^{2}+(1-\alpha)^{2}\rightarrow \alpha(1-\alpha),\ \ 
q^2/2+\epsilon^2\rightarrow -4\epsilon^2.\]

Formulae (\ref{eq8})-(\ref{eq13}) allow to find the nuclear structure function
provided function $\phi$ is known from the evolution equation (\ref{eq7}).
To solve the latter one should fix the initial conditions for the
evolution, that is, the $\phi(y,q,b)$ at the starting point of the
evolution which we denote as $y=0$.

\section{Initial conditions for the evolution}

The initial function $\phi_0(q,b)=\phi(y=0,q,b)$  can be expressed via the
dipole-nucleus cross-section at $y=0$ and fixed $b$:
 \beq
\phi_0(q,b)=\int\frac{d^2r}{2\pi r^2}e^{i{\bf qr}}\Phi_0(r,b)=
\int_0^\infty \frac{dr}{r}{\rm J}_0(qr)\Phi_0(r,b),  \label{eq14}
\eeq
where $\Phi_0(r,b)=\Phi(Y=0,r,b)$.

Staying in the BFKL picture, with a fixed small  strong coupling constant 
$\alpha_s$, at $Y=0$ the scattering of a dipole on the nucleus is
described by the Glauber formula with the dipole-nucleon cross-section
$\sigma(r)$
given by the two-gluon exchange: 
\beq
\Phi_0(r,b)=1-e^{-\frac{1}{2}AT(b)\sigma(r)}.  \label{eq15}
\eeq
Here
\beq
\sigma(r)=g^4\int d^2r^\prime\  G(0,r,r^\prime)\rho_p(r^\prime),
\label{eq16}
\eeq
$\rho_p(r)$ is the colour dipole density in the proton and
$G(0,r,r^\prime)$ is the BFKL Green function at $Y=0$:
\beq
G(0,r,r^\prime)=\frac{rr^\prime}
{8\pi}\frac{r_<}{r_>}\left(1+\ln\frac{r_>}{r_<}\right),\ \ r_{>(<)}={\rm
max(min)} \{r,r^\prime\}.  \label{eq17}
\eeq
$T(b)$ is the nuclear
profile function normalized to 1. 
Obviously the density $\rho_p$ is non-perturbative and not known.
If, as in [8], to simplify the calculations one chooses $\rho_p$ to be
a Yukawa distribution
\beq
\rho_p(r)=a\frac{e^{-\mu r}}{r}\ ,  \label{eq18}
\eeq
then one obtains
\beq
\frac{1}{2}AT(b)\sigma_(r)=B\left[2{\rm C}-1+2\ln \tilde{r}-{\rm
Ei}(-\tilde{r})\left(2+\tilde{r}^2\right)+e^{-\tilde{r}}(1-\tilde{r})\right],
\ \
\tilde{r}=\mu r,  \label{eq19}
\eeq
where C is the Euler constant and
dimensionless $B$ combines the information about the nucleus density,
value of $\alpha_s$ and normalization constant $a$ in (\ref{eq18}).
At $r\to 0$ and $\infty$, (19) behaves as $r^2\ln(1/r)$ and $\ln r$
respectively.

However this sort of initial conditions is only true in the rigorous
BFKL approach with a fixed and very small coupling. In fact, if one
calculates the structure function of the proton using (\ref{eq19}), one finds
a very strong $Q^2$ dependence incompatible with the experimental data.
To be nearer the realistic situation we therefore choose
the dipole-nucleon cross-section $\sigma(r)$ in (\ref{eq15}) to be in agreement
with the DIS data at moderately low $x$, at which one may expect the
start of the evolution according to Eq. (\ref{eq7}). We take 
$x=0.01$ as a starting point for the evolution  and choose $\sigma(r)$
as parametrized by Golec-Biernat and W\"usthoff [17] to reproduce all
DIS data on the proton below $x=0.01$:
\beq
\sigma(r)=\sigma_0\left(1-e^{-\hat{r}^2}\right),\ \ \hat{r}=\beta r,
\label{eq20}
\eeq
with $\sigma_0=29.12$ mb and 
$\beta=0.234$ GeV for 4 quark flavours (with masses $m_u=m_d=m_s=0.14$
GeV and $m_c=1.5$ GeV, which we will use in all our computations).
Nevertheless, these values have been obtained from a fit to experimental
data with $x\leq 0.01$ and we are interested only in
$\sigma(r)$ at $x=0.01$, where the quoted values produce a cross-section
which overestimates the experimental data for $F_2$ of the proton,
while the $Q^2$ evolution at this $x$ is well described.
Thus, to better agree with the experimental data at $x=0.01$
we diminished $\sigma_0$ to a lower value 20.80 mb, keeping the value
of $\beta$; we have verified that with this choice, Eq.
(\ref{eq15}) reasonably reproduces $F_2$ in
nuclei at this value of $x$,
see e.g. [21] for another use of (\ref{eq15}) in this context.
Apart from being in agreement with the experimental data, the cross-section
(\ref{eq20})
has an advantage of allowing to relatively simply Bessel transform the
Glauberized cross-section (\ref{eq15})
to  obtain the initial function $\phi$ in 
the momentum space according to (\ref{eq14}) (see Appendix).

\section {Heavy flavour production off the nucleus}

The total cross-section for the photoproduction of heavy flavour $f$
off the nucleus is given by $\sigma_T$ at $Q^2=0$  with $w_T$
corresponding to the contribution of this  flavour  in the sum  
(\ref{eq12}).
The inclusive cross-section for the production of the heavy quark with
a given transverse momentum $l$ can be obtained from the expression for
the heavy quark density (see [10,22]):
\[
\frac{\partial [xq_f(x,l,b)]}{\partial^2l\partial^2b}
=
\frac{\alpha_sQ^2}{(2\pi)^3}
\int_0^1d\alpha d^2b_1d^2b_2e^{-i{\bf l(b_1-b_2)}}\]\[
\left[\left(\alpha^2+(1-\alpha)^2\right)\epsilon^2\frac{{\bf b_1b_2}}{b_1b_2}
{\rm K}_1(\epsilon b_1){\rm K}_1(\epsilon b_2)+
\left[4Q^2\alpha^2(1-\alpha)^2+m_f^2
\right]{\rm K}_0(\epsilon b_1){\rm K}_0(\epsilon 
b_2) \right]\]\beq
\int\frac{d^2k}{(2\pi)^2}\frac{1}{k^2}
\frac{\partial [xG(x,k,b)]}{\partial^2b\partial k^2}
\left[1+e^{-i{\bf k(b_1-b_2)}}-e^{-i{\bf kb_1}}-e^{i{\bf kb_2}}\right].
\label{eq21}
\eeq
Note that, using (6) and (11) and the fact that $\Phi(Y,0
,b)=0$, the last line of Eq. (21) can be
rewritten as
\[\frac{N_c}{\pi^2g^2}\Big[\Phi(Y,{\bf b_1},b)+
\Phi(Y,{\bf b_2},b)-\Phi(Y,{\bf b_1}-{\bf b_2}, b)\Big];\]
if we neglect the evolution in $Y$ and take into account only the
multiple Glauber rescattering in the nucleus implied by (15),
we recover the result for the quark density in [22] (Eq. (26) in this
reference).
The inclusive cross-section is expressed via the density (21) as
\beq
\frac{d\sigma_f}{d^2l}=\frac{4\pi^2\alpha_{em}}{Q^2}Z_f^2
\int d^2b\frac{\partial [xq_f(x,l,b)]}{\partial^2l\partial^2b}\ .
\label{eq22}
\eeq
Evidently it is finite at $Q^2=0$ with only the transverse  
part surviving in this limit.

The leptoproduction cross-sections can be found once the cross-sections
for the virtual photoproduction are known, which correspond to
(\ref{eq21}) and 
(\ref{eq22}) at non-zero $Q^2$:
\beq
\frac{d\sigma_f(e+A\to e'+Q+X)}{d^2ldQ^2dz}=\frac{\alpha_{em}}{\pi z Q^2} 
\left(1-z+\frac{1}{2}z^2\right)\frac{d\sigma_f(\gamma^*+A\to
Q+X)}{d^2l}  \label{eq23}
\eeq
(see [23] for a discussion on the validity of the equivalent photon
approximation in this kind of computations).
Here $z$ is the fraction of the energy of the incident electron
carried by the virtual photon: $z=q\cdot p/(p_l\cdot p)$,
where $Ap$ is the momentum
of the nucleus and $p_l$ that of the lepton.

It is remarkable that both total and inclusive cross-sections for the
heavy flavour production are directly expressed via 
the gluon density (\ref{eq11}) in our definition.
This perfectly fits the assumed photon-gluon fusion
production mechanism and thus supports our definition of the gluon
density from a pragmatic point of view.

Performing part of the integrations and using Eqs. (\ref{eq10}) and
(\ref{eq11}), we express
the quark density as
\[
\frac{\partial [xq_f(x,l,b)]}{\partial^2l\partial^2b}=
\frac{Q^2N_c}{8\pi^4}\int_0^\infty\frac{dk}{k}h(y,k,b)\]\beq
\int_0^1d\alpha\left\{\left[\alpha^2+(1-\alpha)^2\right]I_T(\epsilon,k)
+\left[4Q^2\alpha^2(1-\alpha)^2+m_f^2\right]I_L(\epsilon,k)\right\},
\label{eq24}
\eeq
where function $h$ is defined by
(\ref{eq10}),
\beq
I_T=
\left(2\epsilon^2+k^2\right)\chi\xi-\epsilon^2\left(
\epsilon^2+l^2+k^2\right)\chi^3-
\epsilon^2\xi^2  \label{eq25}
\eeq
and
\beq
I_L=
\left(\epsilon^2+l^2+k^2\right)\chi^3+\xi^2-2\chi\xi,  \label{eq26}
\eeq
with
\beq
\xi=\frac{1}{\epsilon^2+l^2}\ ,\ \
\chi=\frac{1}{\sqrt{\left(\epsilon^2+(l+k)^2
\right)\left(\epsilon^2+(l-k)^2\right)}}\ .  \label{eq27}
\eeq

These formulae allow to calculate the quark density both at $Q^2=0$ and
at finite $Q^2$ once the gluon density in the nucleus (i.e. function $h$
up to a numerical factor factor) is known from the solution of the 
evolution equation (\ref{eq7}). 

\section{Numerical results}

\begin{figure}[htb]
\begin{center}
\epsfig{file=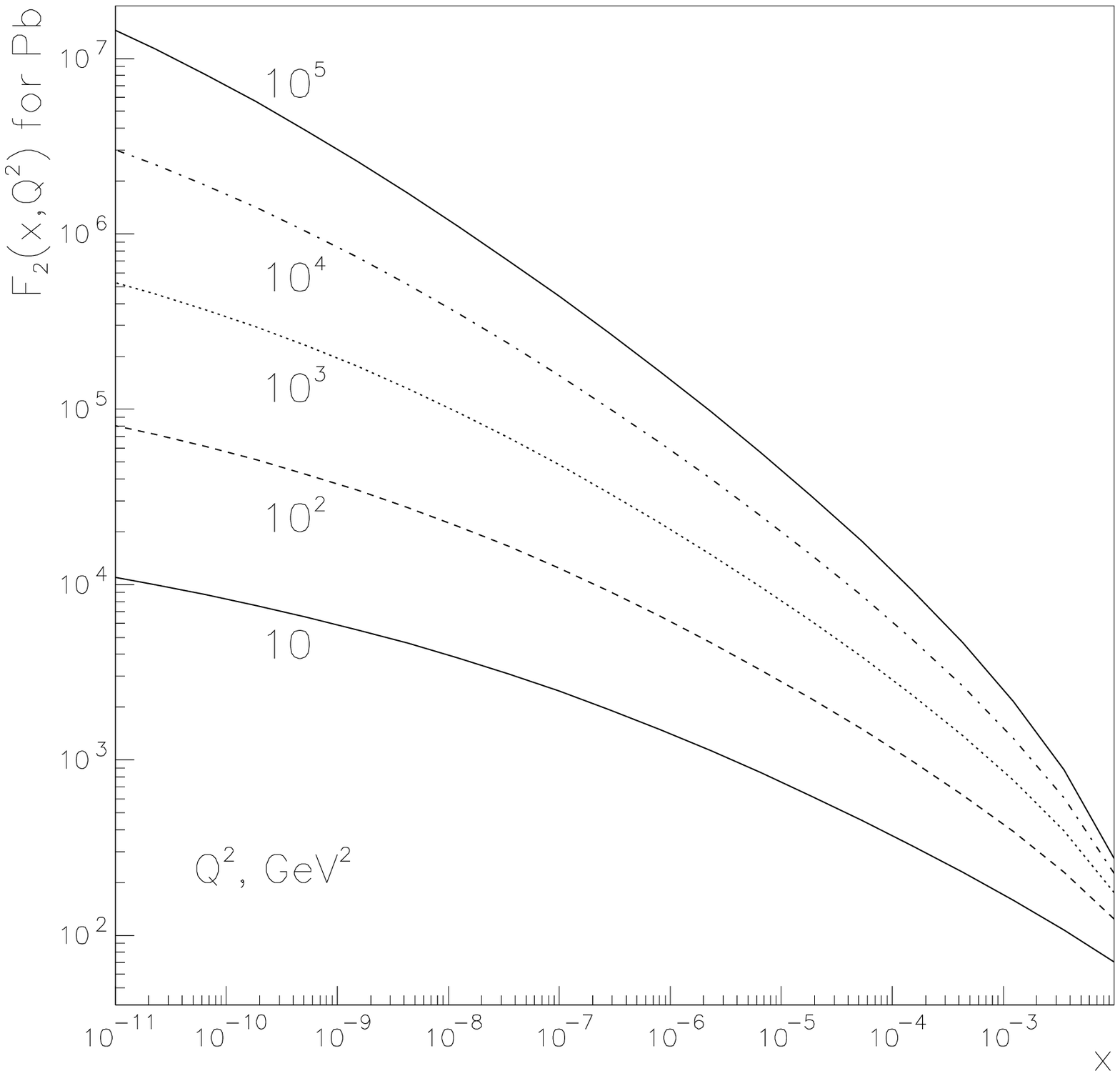,width=15.5cm}
\end{center}
\vskip -1.0cm
\caption{}
\label{fig1}
\end{figure}

\begin{figure}[htb]
\begin{center}
\epsfig{file=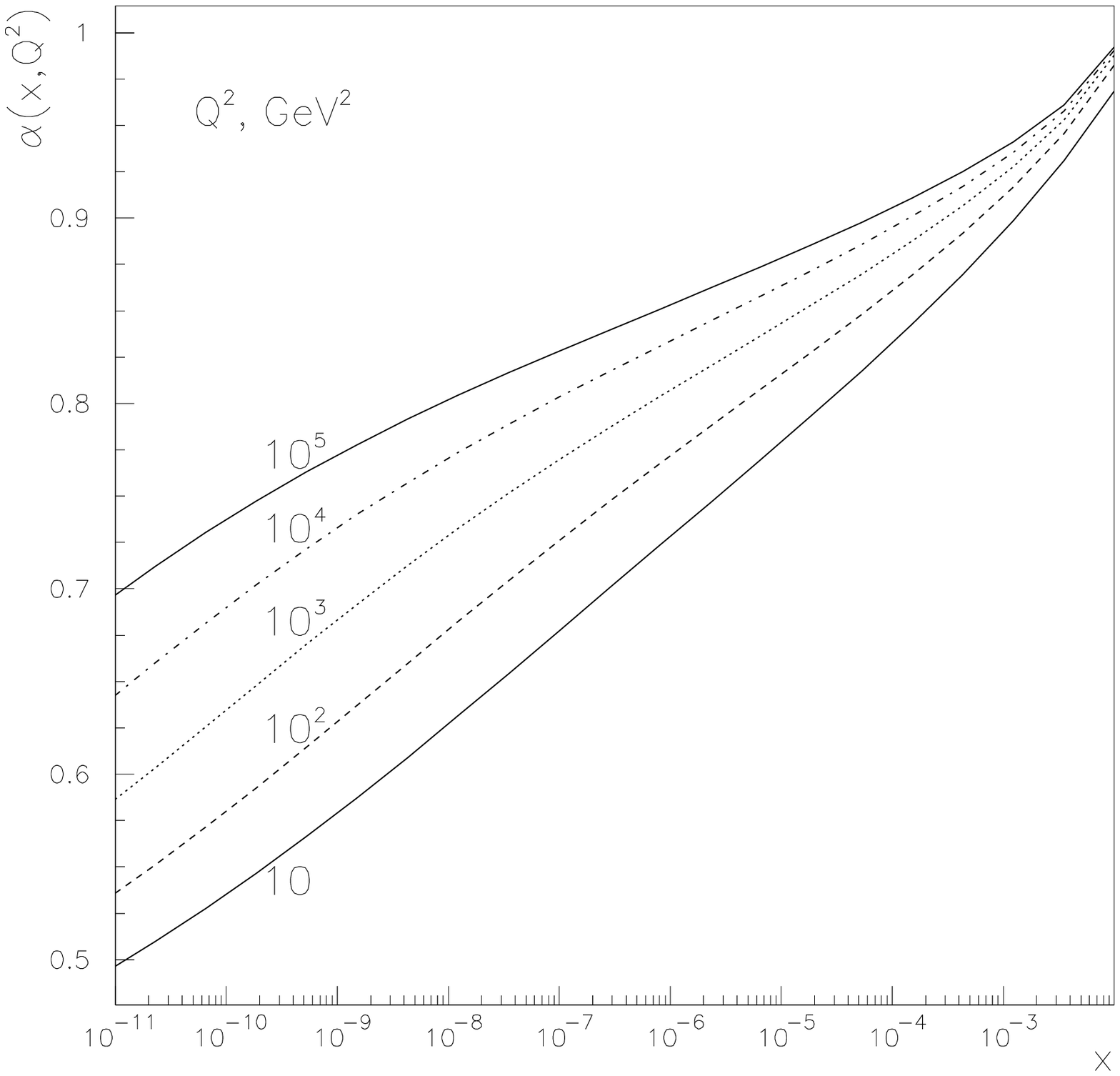,width=15.5cm}
\end{center}
\vskip -1.0cm
\caption{}
\label{fig2}
\end{figure}

Starting from the initial function $\phi_0$ at $x=0.01$ 
described in 
Section 3 we solved numerically the evolution equation (\ref{eq7})
for values of $y$ up to $y_{max}=y_0+4.0$, where $y_0$ is the initial value
of $y$. We have taken $\alpha_s=0.2$, so that $y_0\simeq 0.88$ and
the minimal value of $x$ is $\sim 10^{-11}$.
For the nuclear profile function we use the one corresponding to
the nuclear density
given by a 3-parameter Fermi distribution, with the values of
the parameters taken from [24].
We have taken into account the contribution of charm but
neglected that of bottom, which is relatively small
in the total structure function.
The structure function of Pb in this interval of $x$ and for different
values of $Q^2$ between 10 and $10^5$ (GeV/c)$^2$ is shown in Fig.
\ref{fig1}.
One observes that it grows roughly as  $\ln^2(1/x)$ with $x$,
and as $Q^2$ with $Q^2$. Its $A$-dependence can  be well represented by
a power behaviour
\beq
F_{2A}(x,Q^2)=A^{\alpha(x,Q^2)}F(x,Q^2).  \label{eq28}
\eeq
Power $\alpha$ results  dependent both on $x$ and $Q^2$.
It is presented in Fig. \ref{fig2}.
One observes that at relatively large values 
of $x$ $\alpha$ is close to unity and depends on $Q^2$ very weakly.
However at smaller $x$ the power goes down up to values below 
2/3 naively expected from the nuclear screening. At a given $x$ it  rises 
with increasing $Q^2$,
however the general trend of going down with $1/x$ persists at all
$Q^2$.

\begin{figure}[htb]
\begin{center}
\epsfig{file=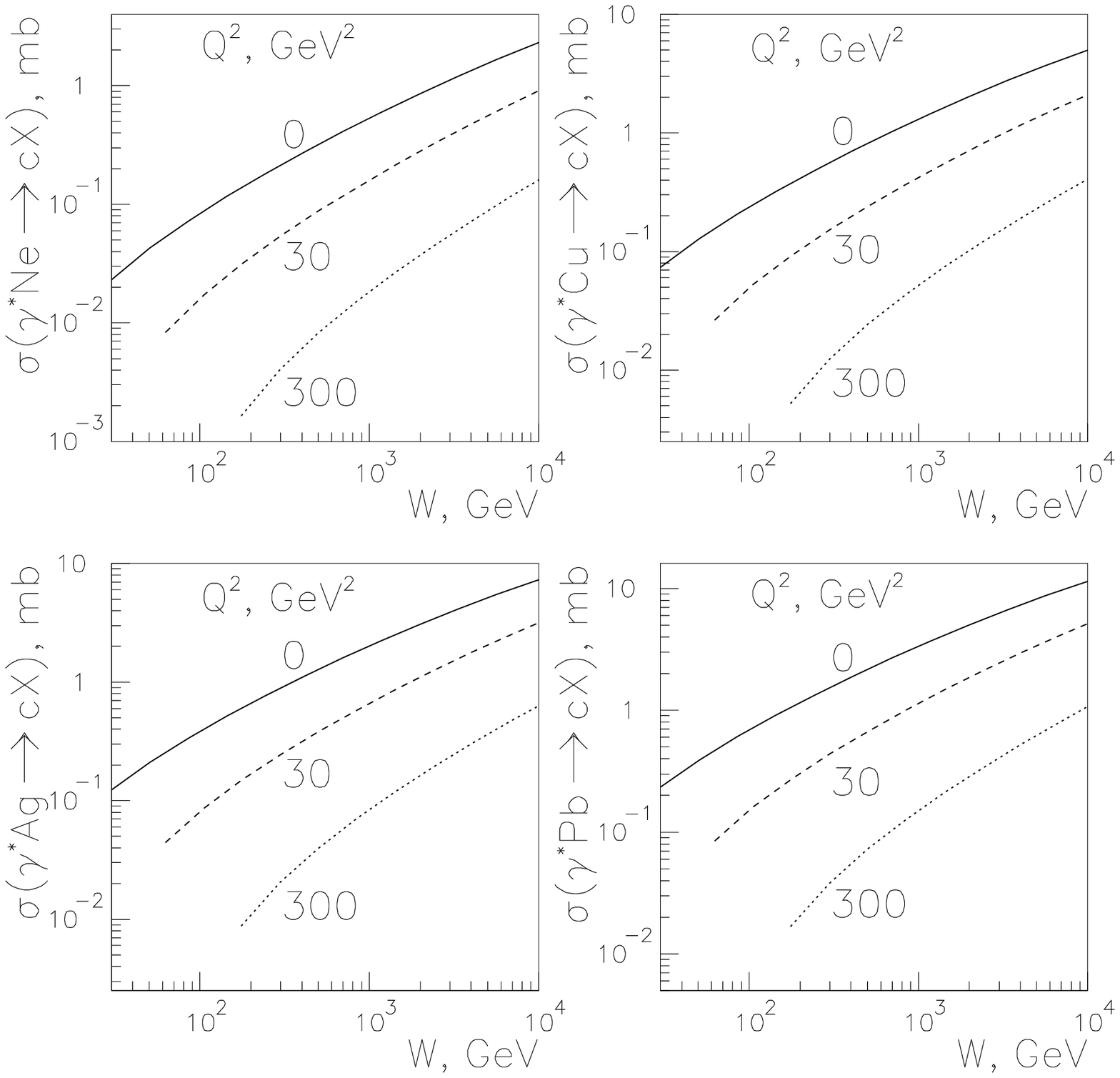,width=15.5cm}
\end{center}
\vskip -1.0cm
\caption{}
\label{fig3}
\end{figure}

\begin{figure}[htb]
\begin{center}
\epsfig{file=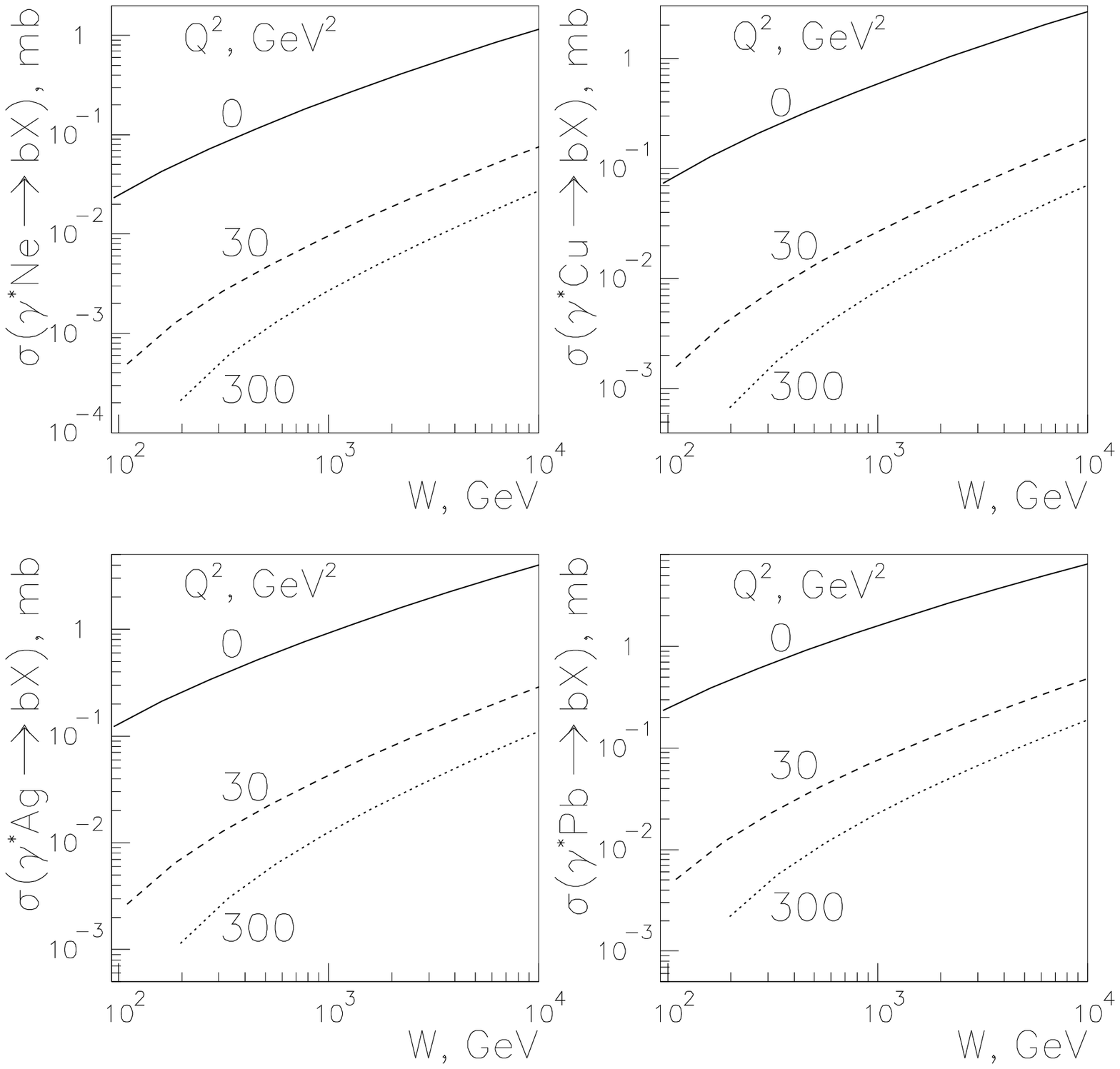,width=15.5cm}
\end{center}
\vskip -1.0cm
\caption{}
\label{fig4}
\end{figure}

Our results for the total cross-section for the 
real and virtual photoproduction of charm and bottom
(with $m_b=4.75$
GeV)
on various nuclei
are presented in Figs. \ref{fig3} and \ref{fig4}.
The cross-sections are shown as  
functions 
of the photon c.m. energy $W$ with respect to a nucleon in the nucleus.
We assumed that
\beq
x=\frac{4m_f^2+Q^2}{W^2+Q^2}\ .  \label{eq29}
\eeq
The minimal value of $W$ corresponds to the maximal value of $x=0.01$
for our solution. Accordingly it rises with $Q^2$, which explains why the
curves for different $Q^2$ start from different $W$.
As expected, the cross-sections rise with increasing $W$ and $1/Q^2$. Their
absolute values are rather big, reaching $\sim 3$ and $\sim 1$ mb for charm and
bottom respectively for Pb at $W=1000$ GeV and $Q^2=0$. The growth with $W$ in
the shown interval $W<10^4$ GeV is rather close to power-like.
However, corresponding  to the behaviour of the structure function
(Fig. \ref{fig1}), at higher $W$ this should change to $\sim \ln^2W$.

\begin{figure}[htb]
\begin{center}
\epsfig{file=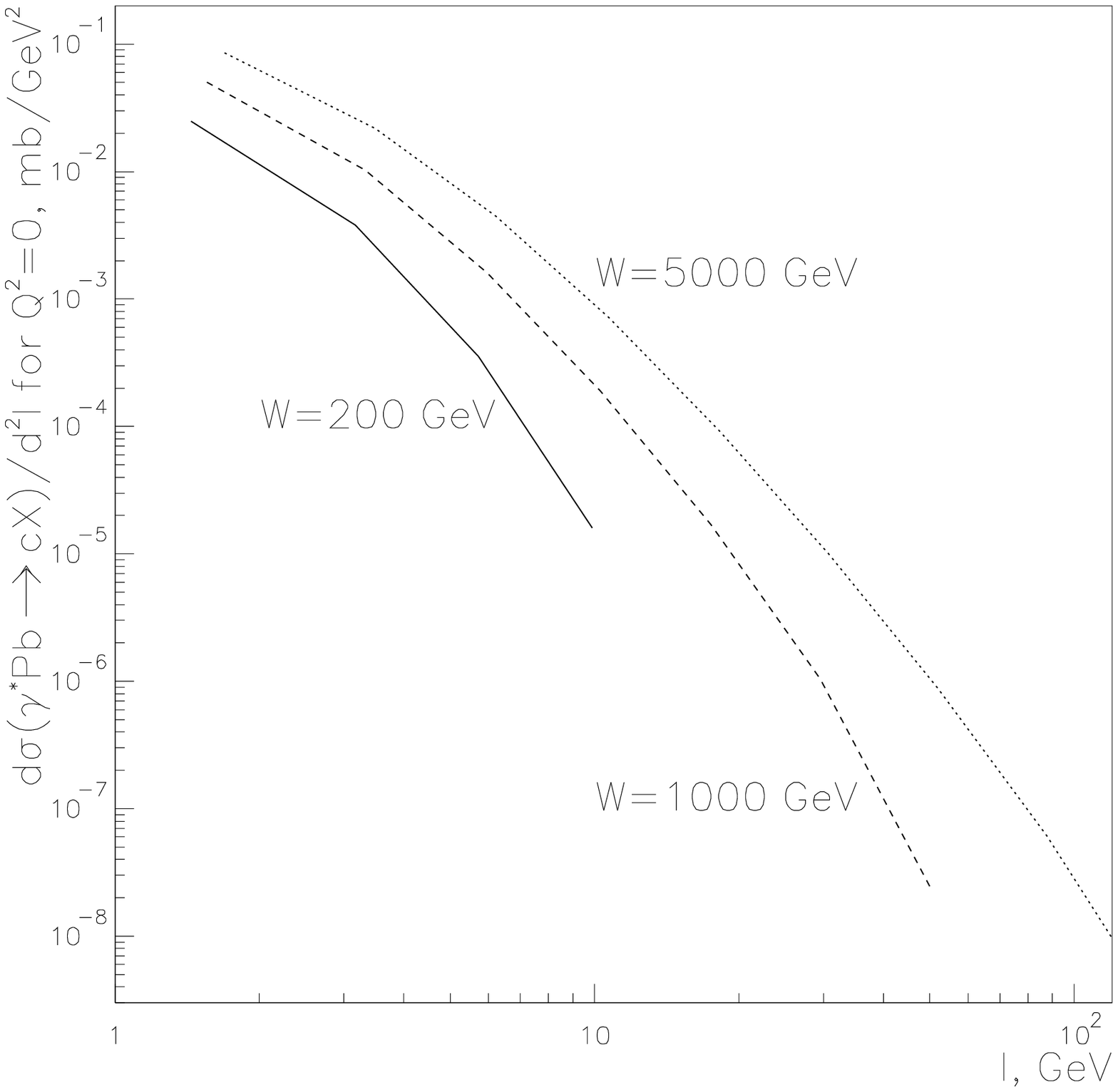,width=15.5cm}
\end{center}
\vskip -1.0cm
\caption{}
\label{fig5}
\end{figure}

\begin{figure}[htb]
\begin{center}
\epsfig{file=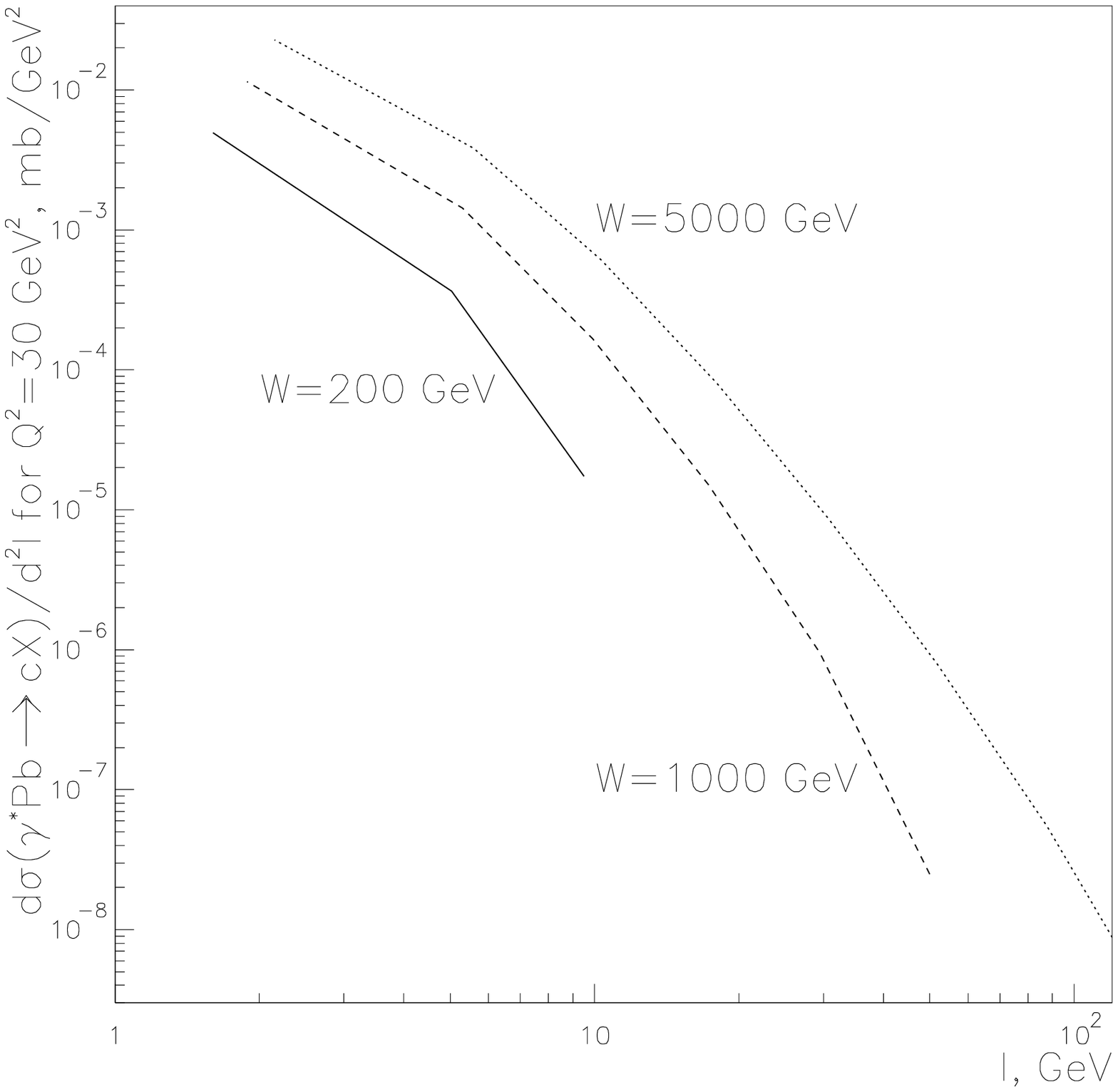,width=15.5cm}
\end{center}
\vskip -1.0cm
\caption{}
\label{fig6}
\end{figure}

In Figs. \ref{fig5} and \ref{fig6}
we show the transverse momentum distribution  
of the produced charmed quark off Pb at $Q^2=0$ and $30$ (GeV/c)$^2$
respectively,
and at $W=200$, 1000 and 5000 GeV (in mb/(GeV/c)$^2$).
Here we assumed that 
\beq
x=\frac{4m_f^2+Q^2+4l^2}{W^2+Q^2}\ .  \label{eq30}
\eeq
Our restriction $x<0.01$, combined with discrete values of $y$
at which the gluon density was calculated, severely limited the number of 
points at small $W$. This explains a somewhat irregular form of the 
curves at $W=200$ GeV. One observes 
that, in accordance with the BFKL kinematics, a relatively large part
of the charmed quarks are produced with momenta of the order or
even larger than the initial large scale $\sqrt{Q^2+4m_f^2}$.

\section{Discussion}

We have calculated the nuclear structure functions at low $x$ which
follow from the fan diagram evolution equation in the perturbative QCD,
with the initial conditions adjusted to the existing experimental data on 
DIS on the proton at $x=0.01$. Knowing the gluon density in the nucleus 
we also calculated the total and inclusive cross-sections for charm and
bottom production at low $x$. The results 
show that the structure functions grow with increasing $1/x$ and $Q^2$ as
$\ln^2(1/x)$ and $Q^2$ respectively. Parametrizing the $A$ dependence as
$A^{\alpha}$ we found that at very small $x$, $\alpha$ is going down to 
values below 1/2. The found transverse distributions of heavy quarks
exhibit a relatively large contribution from the momenta comparable or 
even larger than the natural scale $\sqrt{4m_f^2+Q^2}$.

Our results have been obtained in the pure BFKL kinematical regime 
without any additional cuts. However they are based on the approximation 
of large $N_c$. Therefore one should not expect that they remain valid 
at extremely small $x$ when the $1/N_c^2$ corrections growing with $1/x$
may change the found behaviour.
 
We can compare our results for the structure functions with the recently
obtained in [12]. Although the general trend is similar, our structure 
functions grow with $1/x$ considerably faster. In particular, 
for Au at $Q^2=100$ (GeV/c)$^2$ and $x=10^{-7}$, and also taking $\alpha_s=0.25$
as in [12], we obtain  $F_{2A}/A\sim 140$ as compared to $\sim 20$ in [12].
The difference may come partly from a different initial condition,
which in [12] was taken in a simplified form (with $\sigma(r)$ roughly
proportional to $r^2$) and partly from the fact that in [12] the 
asymptotical form for the photon dipole distributions (\ref{eq3}) and
(\ref{eq4})
at large $Qr$ was used. The latter approximation seems of dubious validity
to us, since in fact values $Qr\sim 1$ give the bulk of the contribution 
to the structure function. Besides, in [12] the integration over the dipole
dimension $r$ was cut at both small and large values, which obviously
introduced a dependence on the cut-off parameters.

\section*{Acknowledgments}

M. A. B.  acknowledges financial support by Secretar\'{\i}a de Estado de
Educaci\'on y Universidades of Spain, and also by a grant of RFFI of Russia. 
 N. A. acknowledges financial support by CICYT of Spain
under contract AEN99-0589-C02, and by Universidad de C\'ordoba.
The authors are thankful to Profs. C. Merino,
C. Pajares and G. Parente for
attention to this work and helpful discussions.

\section{Appendix. The initial function with the 
Golec-Biernat-W\"usthoff
dipole cross-section}

Using Eqs. (\ref{eq14}), (\ref{eq15}) and (\ref{eq20}) and
passing to the integration over $x=\beta r$ we write the initial function
$\phi_0(q,b)$ as an integral
\beq
\phi_0(q,b)=\int_0^{\infty}\frac{dx}{x}{\rm J}_0(qx/\beta)\left(
1-e^{-B\left[1-e^{-x^2}\right]}\right). \label{eq31}
\eeq
The dependence on $b$ is contained in the dimensionless factor $B$:
\beq
B=\frac{1}{2}AT(b)\sigma_0.  \label{eq32}
\eeq
In the following this dependence will be suppressed. We will also denote
$z=q/\beta$.

We present the exponential as a power series in its argument:
\beq
1-e^{-B\left[1-e^{-x^2}\right]}
=-\sum_{n=1}^\infty\frac{(-B)^{n}}{n!}\left(1-e^{-x^2}\right)^n,
\label{eq33}
\eeq
so that
\beq
\phi_0(q)=-\sum_{n=1}^\infty\frac{(-B)^{n}}{n!}I_n(z),  \label{eq34}
\eeq
where
\beq
I_n(z)=\int_0^{\infty}\frac{dx}{x}{\rm J}_0(xz)
\left(1-e^{-x^2}\right)^n.  \label{eq35}
\eeq

The integral $I_n$ can be presented as
\beq
I_n(z)=\sum_{k=0}^nC_n^k(-1)^k
\int_0^{\infty}\frac{dx}{x}{\rm J}_0(xz)
e^{-kx^2}=
-\sum_{k=0}^nC_n^k(-1)^k
\int_0^{\infty}\frac{dx}{x}{\rm J}_0(xz)
\left(1-e^{-kx^2}\right).  \label{eq36}
\eeq

The last integral in $x$ can be found analytically. Indeed one has
(see [25])
\beq
\int _0^{\infty}dx\ x^{\mu}e^{-\alpha x^2}{\rm J}_\nu(z x)=
\frac{z^{\nu}\Gamma((\mu+\nu+1)/2)}{2^{\nu+1}\alpha^{(\mu+\nu+1)/2}
\Gamma(\nu+1)}{_1F_1}\left(\frac{\mu+\nu+1}{2};\nu+1;-\frac{z^2}{4\alpha}\right).
\label{eq37}
\eeq
Taking $\nu=0$ one has
\beq
\int _0^{\infty}dx\ x^{\mu}e^{-\alpha x^2}{\rm J}_0(z x)=
\frac{\Gamma((\mu+1)/2)}{2\alpha^{(\mu+1)/2}}
{_1F_1}\left(\frac{\mu+1}{2};1;-\frac{z^2}{4\alpha}\right).  \label{eq38}
\eeq

We take the difference of two integrals (\ref{eq38}) with different
$\alpha$'s,
\[
\int _0^{\infty}dx\ x^{\mu}\left(e^{-\gamma x^2}-e^{-\alpha x^2}\right)
{\rm J}_0(z x)=\]\beq
\frac{1}{2} \Gamma((\mu+1)/2)
\left[{\gamma^{-(\mu+1)/2}}
{_1F_1}\left(\frac{\mu+1}{2};1;-\frac{z^2}{4\gamma}\right)-
{\alpha^{-(\mu+1)/2}}
{_1F_1}\left(\frac{\mu+1}{2};1;-\frac{z^2}{4\alpha}\right)\right].  \label{eq39}
\eeq
The result is obviously finite at $\mu=-1$. To find it we put
\beq
\mu+1=2\Delta\to 0.  \label{eq40}
\eeq
In this limit,
\beq
\gamma^{-\Delta}\simeq 1-\Delta\ln\gamma,\ \ 
\alpha^{-\Delta}\simeq 1-\Delta\ln\alpha,\ \ 
\Gamma(\Delta)\simeq 1/\Delta  \label{eq41}
\eeq
and
\beq
{_1F_1}(\Delta;1;t)\simeq 1+\Delta\Big({\rm Ei}(t)-C-\ln(-t)\Big),\ \
t<0.  \label{eq42}
\eeq
Putting this into (\ref{eq39}) we find, in the limit $\Delta\to 0$,
\beq
\int _0^{\infty}\frac{dx}{x}\left(e^{-\gamma x^2}-e^{-\alpha x^2}\right)
{\rm J}_0(z x)=\frac{1}{2}\left[{\rm Ei}\left(-\frac{z^2}{4\gamma}\right)-
{\rm Ei}\left(-\frac{z^2}{4\alpha}\right)\right].  \label{eq43}
\eeq
Taking here $\gamma\to 0$ we find the desired integral:
\beq
\int _0^{\infty}\frac{dx}{x}\left(1-e^{-\alpha x^2}\right)
{\rm J}_0(z x)=-\frac{1}{2}
\ {\rm Ei}\left(-\frac{z^2}{4\alpha}\right).  \label{eq44}
\eeq

With the help of this formula one can find the integrals $I_n(z)$ as
a finite sum of exponential integrals of different arguments, Eq.
(\ref{eq36}).
Putting this into (\ref{eq34}) gives the initial function $\phi_0$. For
realistic values of the parameter $B\leq 3$ the convergence of the series
in $n$ is very fast, so that $\phi_0$ can be calculated with very high
accuracy at all $q$.

\section*{References}
%

\noindent [1] A. H. Mueller, Nucl. Phys. {\bf B415} (1994) 373;
A. H. Mueller and B. Patel, Nucl. Phys. {\bf B425} (1994) 471.

\noindent [2] N. N. Nikolaev and B. G. Zakharov, Z. Phys. {\bf C64} (1994)
631.

\noindent [3] L. V. Gribov, E. M. Levin and
M. G. Ryskin, Phys. Rept. {\bf
100} (1983) 1.

\noindent [4] A. H. Mueller and J. Qiu, Nucl. Phys. {\bf B268} (1986) 427.

\noindent [5] A. L. Ayala Filho, M. B. Gay Ducati and E. M. Levin,
Nucl. Phys. {\bf B493} (1997) 305.

\noindent [6] I. I. Balitsky, hep-ph/9706411; Nucl. Phys. {\bf B463} 
(1996) 99.

\noindent [7] Yu. V. Kovchegov, Phys. Rev. {\bf D60} (1999) 034008;
{\bf D61} (2000) 074018.

\noindent [8] M. A. Braun, Eur. Phys. J. {\bf C16} (2000) 337.

\noindent [9] E. Iancu, A. Leonidov and L. McLerran, hep-ph/0011241;
Phys. Lett. {\bf B510} (2001) 133; E. Iancu and L. McLerran,
Phys. Lett. {\bf B510} (2001) 145.

\noindent [10] N. Armesto and M. A. Braun, Eur. Phys. J. {\bf C20}
(2001) 517.

\noindent [11] E. M. Levin and K. Tuchin, Nucl. Phys. {\bf B573} (2000) 833;
hep-ph/0101275.

\noindent [12] E. M. Levin and M. Lublinsky, hep-ph/0104108.

\noindent [13] M. A. Kimber, J. Kwieci\'nski and A. D. Martin,
Phys. Lett. {\bf B508} (2001) 58.

\noindent [14] M. Lublinsky,
hep-ph/0106112.

\noindent [15] M. A. Braun, hep-ph/0101070.

\noindent [16] H. Abramowicz {\it et al.}, {\it
TESLA Technical Design Report, Part VI, Chapter 2}, Eds. R. Klanner, U.
Katz, M. Klein and A. Levy.

\noindent [17] K. Golec-Biernat and M. W\"usthoff, Phys. Rev. {\bf D59} 
(1999) 014017; {\bf D60} (1999) 114023.
 
\noindent [18] A. Capella, E. G. Ferreiro, A. B. Kaidalov and C. A.
Salgado, Nucl. Phys. {\bf B593} (2001) 336; Phys. Rev. {\bf D63} (2001) 054010.
 
\noindent [19] N. Armesto and C. A. Salgado, hep-ph/0011352.

\noindent [20] M. A. Braun, hep-ph/0010041.

\noindent [21] N. Armesto and M. A. Braun,
Z. Phys. {\bf C75} (1997) 709.

\noindent [22] A. H. Mueller, Nucl. Phys. {\bf B558} (1999) 285.

\noindent [23] S. Frixione, M. L. Mangano,
P. Nason and G. Ridolfi,
Phys. Lett. {\bf B319} (1993) 339.

\noindent [24] C. W. De Jager, H. De Vries and C. De Vries,
Atom. Data Nucl. Data Tabl. {\bf 14} (1974) 479.

\noindent [25] I. S. Gradshteyn and I. M. Ryzhik, {\it Table of
Integrals, Series and Products}, Academic Press 1994.

\section*{Figure captions}

\noindent{\bf Fig. 1:}
The structure function $F_2$ of Pb as a function of $x$ at different
$Q^2$.

\noindent{\bf Fig. 2:}
The power $\alpha$ of the $A$-dependence of the nuclear structure functions
as a function of $x$ at different $Q^2$.

\noindent{\bf Fig. 3:}
The total cross-sections for real and virtual photoproduction of charm
on Pb (lower right plot), Ag (lower left plot), Cu (upper right plot)
and Ne (upper left plot).

\noindent{\bf Fig. 4:}
The total cross-sections for real and virtual photoproduction of bottom
on Pb (lower right plot), Ag (lower left plot), Cu (upper right plot)
and Ne (upper left plot).

\noindent{\bf Fig. 5:}
The inclusive cross-section $d\sigma/d^2l$ for the
real photoproduction of charmed quarks on Pb, as a function of the
photon-nucleon
c.m. energy $W$.

\noindent{\bf Fig. 6:}
The inclusive cross-section $d\sigma/d^2l$ for the
virtual photoproduction at $Q^2=30$ (GeV/c)$^2$ of charmed quarks on Pb, as a 
function of the photon-nucleon c.m. energy $W$.

\newpage  

\newpage  

\newpage  

\newpage  

\newpage  

\newpage  

\end{document}